# Accounts, Accountability and Agency for Safe and Ethical AI

**Rob Procter**
Department of Computer Science
Warwick University & Alan Turing Institute for Data Science and AI
Coventry
CV4 7AL, UK
rob.procter@warwick.ac.uk

**Mark Rouncefield**
School of Computing and Communications
Lancaster University
Lancaster
LA1 4YW, UK
m.rouncefield@lancaster.ac.uk

**Peter Tolmie**
Information Systems and New Media
University of Siegen
57068 Siegen
peter.tolmie@uni-siegen.de

**ABSTRACT**

We examine the problem of explainable AI (xAI) and explore what delivering xAI means in practice, particularly in contexts that involve formal or informal and ad-hoc collaboration where agency and accountability in decision-making are achieved and sustained interactionally. We use an example from an earlier study of collaborative decision-making in screening mammography and the difficulties users faced when trying to interpret the behavior of an AI tool to illustrate the challenges of delivering usable and effective xAI. We conclude by setting out a study programme for future research to help advance our understanding of xAI requirements for safe and ethical AI.

# INTRODUCTION

Explainable AI (xAI), i.e., AI's capacity for transparency and interpretability, is now recognised as a key requirement for the effective and safe application of AI tools [1]. As a consequence, much effort is now being devoted to developing ways to address the limited degree to which current machine learning techniques are able to satisfy this requirement [2-5]. This has led to the identification of a number of promising techniques for xAI [6], such as contrastive explanations, which are counter-factuals that may be applied globally (i.e., techniques that attempt to explain the model as a whole) or locally (i.e. techniques that attempt to explain the model's behaviour for a specific input). Recently, we have seen evidence of a 'turn to the social' in xAI:

> "Explanations are social, insofar as they involve an interaction between one or more explainers and explainees…" [5]

We therefore need to ensure that research in xAI is able to lead to methods for producing explanations that reflect this understanding. Below, we unpack what meeting this requirement of xAI entails through the lenses of *agency* and *accountability*.

# UNDERSTANDING REQUIREMENTS FOR xAI

Designing explainability into AI tools is essential if they are to be trusted [7] and if their users are to be able to exercise *agency* when making decisions, whether they be professional [7-8] or lay users [9]. In other words, AI tools must be *accountable* to users for the ways in which they behave [10-11]. There are clearly moral, ethical and legal aspects to this. Of particular relevance is Brey's notion of the 'disclosive' or embedded ethics of computer systems [12-13]. Fulfilling these requirements is challenging for a number of reasons.

First is the problem of designing accounts – "computational representations which systems continuously offer of their own behaviour and activity" [14] – so that they will be understood by the user in a range of different circumstances and thus satisfy a requirement of being "a resource for improvised and contextualised action" (Ibid), while preserving the user's *agency* as a decision-maker.

Secondly, in the workplace, agency is a property that is distributed and performed collaboratively. Thus, it depends on participants being able to make sense of the actions of others and to make their own actions understood by others [8, 15-17]. Hence, accounts furnished by AI tools may also serve as resources for collaboration and their design should not only satisfy the needs of individual users but also reflect how (formally and informally) collaboration is done. In other words, they must afford the articulation work [18] on which participants may rely for maintaining their *mutual accountability* [15, 19-20]. This is a feature of what Gray et al. [21] term 'corollary work': "emergent socio-material practices of place, time, productivity and self-expression".

Thirdly, whether by design or unintentionally, new technologies in the workplace are likely to result in changes in work practices [22] and the implications of these changes for AI tool accounts for agency and mutual accountability must be understood if tasks are to be performed in ways that are timely and dependable [23-24].

Finally, and not least, these accounts may assume a formal role in that, by being recordable, they may be seen as furnishing ways to further organisational goals of holding members to account for their past actions [25]. So, the implications of this also need to be investigated and understood.

**AN EXAMPLE: READING MAMMOGRAMS IN THE UK BREAST SCREENING PROGRAMME**

Healthcare is a particularly active area for the adoption of AI. Recently, for example, trials of new AI tools for screening mammography have been conducted in the UK [26]. A previous study we conducted of radiologists reading mammograms in the UK breast screening programme may help to understand what the requirements for xAI might be in this context [27-28].

In the UK breast screening programme, mammograms are read by two radiologists and the recall/no recall decision is made on the basis of these two (semi-)independent assessments (ref). We see in Figure 1 a radiologist examining mammograms (2 views: 'oblique' and 'CC'). As accounts, mammograms may seem of limited value, but the radiologist is able to work up an account by, e.g.: (a) comparing features across the 2 views; (b) using a magnifying glass; and (c) measuring features using their hands [29].

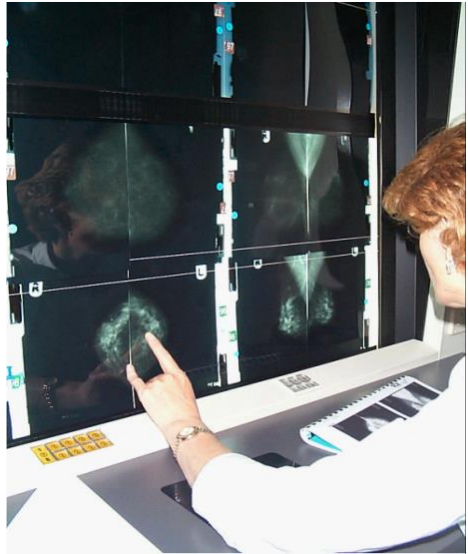

**Figure 1: Radiologist examining a mammogram, showing use of fingers to measure a feature**.

**Figure 2: Example of a UK Breast Screening report form.**

Figure 2 shows the screening form that is used to record the comments and decisions the participants make during the process. This begins when the mammograms are taken by the radiographer, who records information about, e.g., exposure but also information gleaned from the woman (cyst; moles; Pain L breast and arm, GP thinks is muscular) that the radiographer adds in order to be *accountable*. Comments are indexically tied to the mammograms through marking the simple schematics. Similarly, the 1st radiologist doesn't just record recall/no recall but adds a comment (new), which is then available to the 2nd radiologist, who adds a final comment (BT I think, HRT related). The combination of films and form provides accountability for the people involved and thus for the *whole process*.

Figure 3 shows an early prototype of a computer-aided mammography tool. At the top are the original mammograms and at the bottom are displays showing prompts for areas the system (CADe) thinks are suspicious. We studied radiologists using CADe in order to understand how they made sense of the prompts and how the prompts influenced their decisions [30]. Our studies showed that reading mammograms has collaborative aspects, achieved through making available features that are professionally relevant. This is what Goodwin [31] terms 'professional vision', a 'way of seeing', a technique for making relevant features available and accountable, and with which any technologies need to interact and accommodate. However, evidently, the machine knows nothing of what it is to be a competent, professional reader and is (currently) unable to 'explain' its prompts and detections; instead the reader must 'repair' what the machine shows, thereby making it 'accountable'.

This then is the problem of xAI. As CADe was not capable of providing an account for its recall/no recall prompts, the radiologists needed to come up with one based on their accumulated observations of its behaviour. Though useful at times, such ad-hoc explanations may also be unreliable and misleading, underlining the importance of the problem that xAI aims to address.

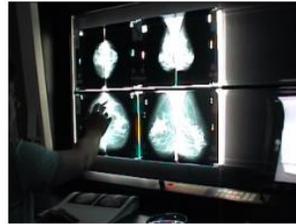 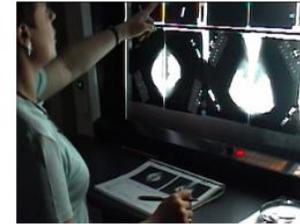

CADe's got a whole row of markers up here… but they're all on innocent things… nothing to worry about there…

I've seen this before… the computer's marked something that I think is artefact on that side… it's often along the edge of the film.

**Figure 3: Radiologists interpreting the behaviour of CADe.**

**A STUDY PROGRAMME FOR xAI**
We plan to investigate, through a series of case studies in different work settings, the requirements for xAI in context. The findings will be used to inform recommendations for design [32], including identification of design patterns [33-34] for accountability that have potential to be reusable in future xAI design. The case studies will also be used to explore co-production methodologies that enable users to have agency in the design and development process [35-37] and to ensure that organisational learning is properly supported over time [38].

We will use a combination of interviews and ethnographic studies to explore work practices in a range of current or potential settings for the application of AI tools. Following the 'turn to the social' [38-42], ethnography has become a well-established tool for IT systems requirements gathering. Its value lies in its capacity to recover the 'real world' aspects of a setting, identifying the exceptions, contradictions and contingencies of activities that do not necessarily figure in their more formal representations.

Case studies will be drawn from actual or potential application domains for AI tools, including: healthcare: digital pathology; assistive technologies; and air traffic control. Regarding the latter, Bentley et al.'s [43] study of air traffic control is a seminal example of the value of ethnography for IT systems requirements gathering, revealing important but sometimes neglected ways in which social affordances of technologies contribute to dependable conduct and decision-making.

ACKNOWLEDGMENTS
We would like to acknowledge the support of the UK Alan Turing Institute for Data Science and AI.